\documentclass[twoside]{article}
\usepackage{fleqn,espcrc2}
\usepackage{graphicx}

\newcommand{\AmS}{{\protect\the\textfont2
  A\kern-.1667em\lower.5ex\hbox{M}\kern-.125emS}}

\title{Higgs Production via Gluon-Induced Weak Boson Fusion}

\author{J. Vollinga\address{Nikhef Theory Group, Kruislaan 409, 1098 SJ Amsterdam, The Netherlands}}
       
\begin{document}

\begin{abstract}
We present a calculation that allows for an estimation of the order $\alpha^3\alpha_s^2$
contributions to the Higgs production in the weak boson fusion channel. A possible deterioration of
this important channel for the Higgs discoveries at the LHC can be ruled out by this calculation
due to the small remaining cross section after the weak boson cuts.
\end{abstract}

\maketitle

\section{Introduction}

One of the primary goals of the CERN Large Hadron Collider (LHC) that will go into full operation
soon is to illuminate the electro-weak symmetry breaking mechanism. The favoured mechanism --- the
Higgs mechanism --- predicts the existence of a Higgs boson. The Higgs boson will be produced at the
LHC mainly through gluon fusion which has the largest cross section there. While for Higgs masses of
$m_H=160$ GeV or higher the Higgs decays almost exclusively into weak boson pairs giving a well
identifiable signature in the detectors, for lower Higgs masses the decay is mainly into $b\bar{b}$
pairs. This decay can hardly be distinguished from other QCD background processes so that for the
Higgs production by gluon fusion the searches can only rely on rare Higgs decays into photon pairs.

The second greatest production channel of the Higgs boson is the weak boson fusion (WBF) and can
reach up to 20\% of the leading order gluon fusion cross section. This channel has a distinctive
signature. Since in leading order no colored particles are exchanged between the quarks the signal
consists of two forward jets with almost no hadronic activity in the central region. This signature
allows for a good reconstruction of the signal events and makes this channel important for Higgs
searches especially for small Higgs masses \cite{Rain97}.

The LO and NLO cross sections of the WBF Higgs production have been calculated
\cite{Rain97}\cite{Han92}\cite{Djou99}.  The NLO correction gives an increase in rate of about 10\%
while maintaining the good decay signature. Most one-loop diagrams in NLO order are of the DIS type,
i.e. being vertex corrections.  Some NLO diagrams do have a gluon exchange between the quark lines,
but these diagrams contribute only in interference with crossed LO diagrams. This t--u--channel
interference is very small and can therefore be neglected.  Effectively, the NLO contribution has no
t--channel color exchange and so no additional hadronic activity in the central region can be found.

The situation for the NNLO contributions is different. Here one can expect an alteration of the
kinematic distributions that might destroy the virtues of the WBF channel, because sizable
interferences between two one-loop diagrams as well as between two-loop diagrams and t--channel tree
diagrams are possible. Usually, NNLO contributions are expected to be small, but in this case this
might not be true due to the new possibility of having gluons in the initial state and considering
the high gluon luminosity at the LHC. This asks for an investigation to answer the question whether
the discovery potential of the WBF channel is reduced.

In the following we will describe a calculation we have done to estimate the effect of the NNLO
contribution on the WBF channel Higgs discovery potential. A more detailed presentation of the
calculation can be found in \cite{Harl08}.

\section{Scope of the calculation}

\begin{figure}[htb]
	\begin{center}
		\begin{tabular}{ccc}
			\includegraphics[width=2.1cm]{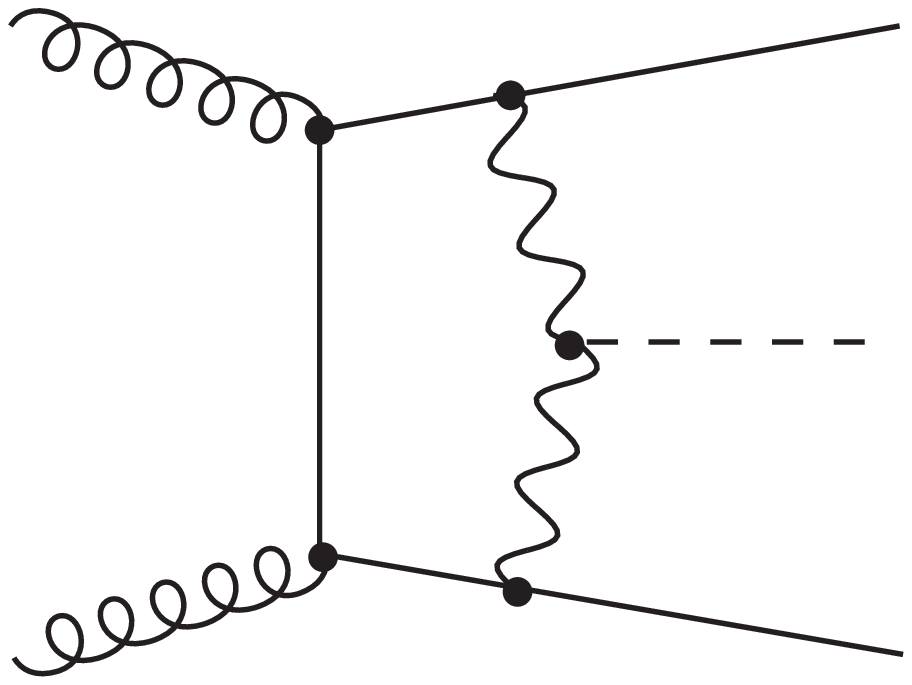} &
			\includegraphics[width=2.1cm]{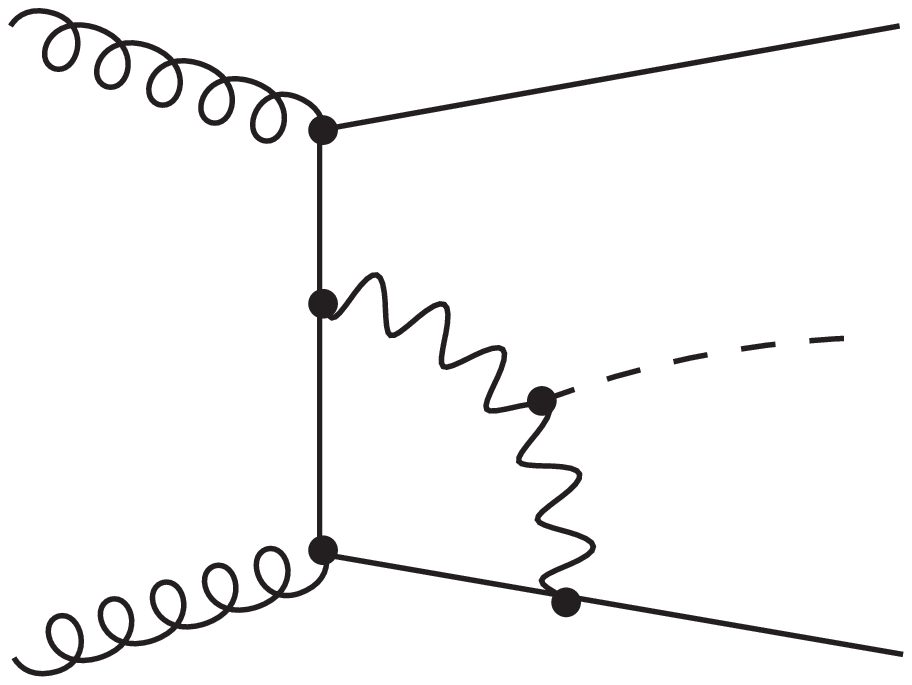} &
			\includegraphics[width=2.1cm]{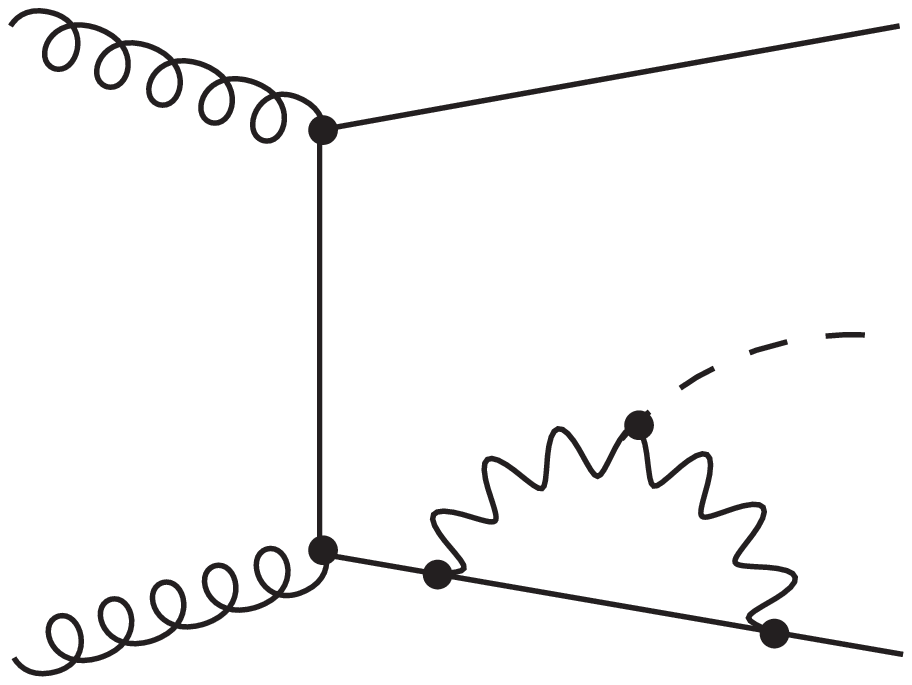} \\
			\includegraphics[width=2.1cm]{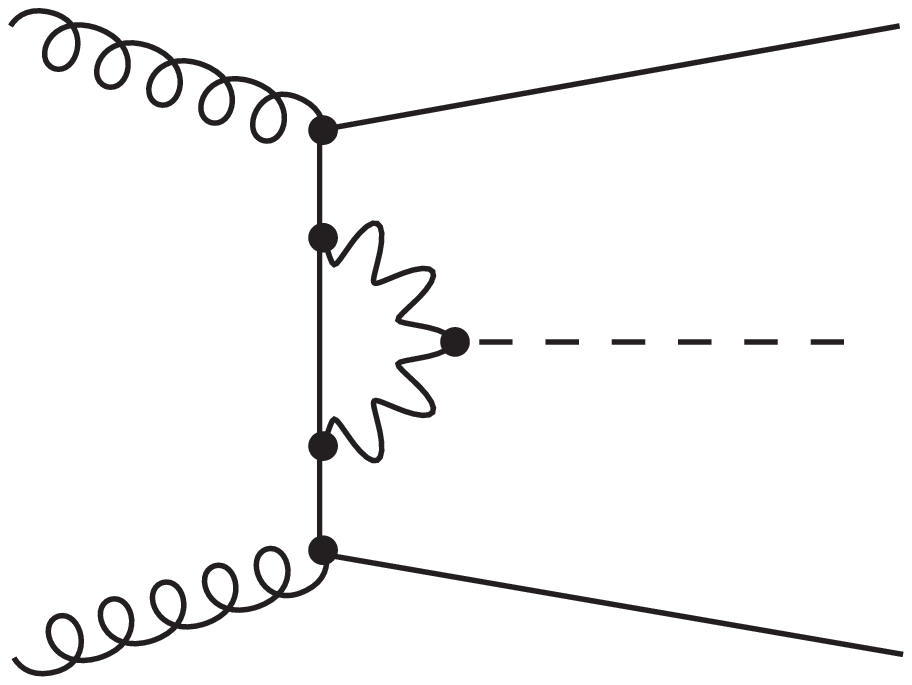} &
			\includegraphics[width=2.1cm]{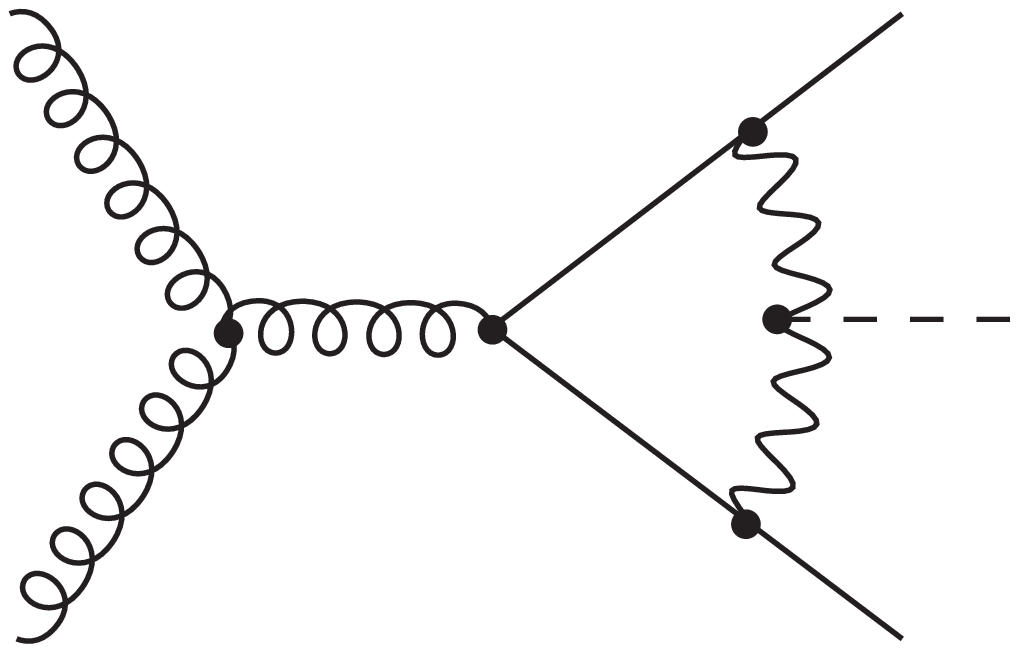} &
			\includegraphics[width=2.1cm]{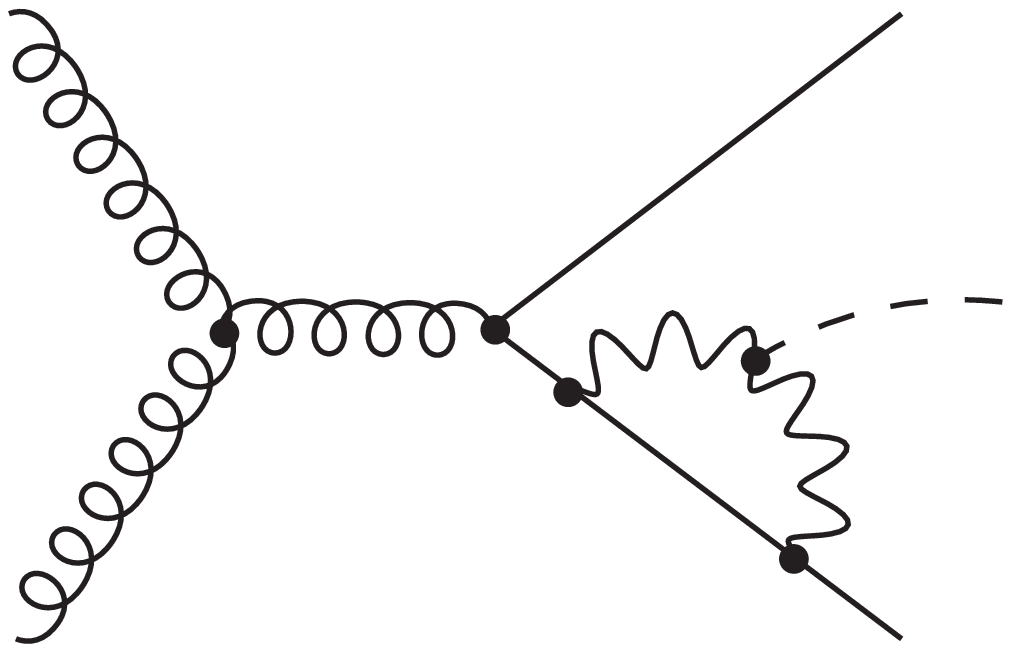}
		\end{tabular}
		\vspace*{-5mm}
		\caption[]{\label{fig:diagrams}\sloppy
		Representative selection of diagrams that were calculated. Diagrams with $qg$,
		$\bar{q}g$, or $q\bar{q}$ initial states can be derived by simple crossing.}
	\end{center}
\end{figure}

We want to calculate the NNLO QCD corrections to the WBF process. The order of the calculation is
formally $\alpha^3\alpha_s^2$. We are not aiming for a full NNLO calculation but try to estimate
possible NNLO effects on the WBF signature and their impact on the Higgs searches. Therefore we
remove as many diagrams as possible from the calculation to arrive at a sufficient and gauge
invariant subset.

We do not consider DIS-like contributions where the gluons are only involved in $q\bar{q}V$ vertex
corrections. The size of these contributions can be estimated from \cite{Zijl92} and no deterioration of
the WBF signature is to be expected.

Double gluon exchange between the quark lines gives rise to the most difficult diagrams to calculate
having two loops. If the gluons do not form a color singlet then the interference with the
tree-level diagram is in the t--u--channel and can be neglected. On the other hand, if the gluons
form a color singlet then we don't expect additional hadronic activity in the central region, so the
WBF signature is not altered. Therefore we do ignore these double gluon exchange diagrams and
together with them also their double real radiation counterparts.

Diagrams in which the $Z$ boson becomes resonant are also neglected. Those diagrams have the Higgs
radiated off the $Z$ and the $Z$ then decays into a $q\bar{q}$ pair. This set of diagrams belongs to
the Higgsstrahlung process $q\bar{q}\rightarrow HZ$ and is to be treated within the appropriate
analyses for Higgsstrahlung.

To simplify the calculation further we take all quarks except the top quark to be massless. Then the
Higgs can only couple to the weak bosons or the top quark and we have to consider only up-type,
down-type, and top quarks as fermions in the diagrams.

The remaining diagrams have the characteristic feature of possessing only a single quark line. Fig.
\ref{fig:diagrams} shows some example diagrams with gluon pairs in the initial state. Other diagrams
with $qg$, $\bar{q}g$, and $q\bar{q}$ initial states can be derived from them by crossing. We
also neglect the diagrams with top quarks in the initial state due to the tiny available
phase space.

\section{Calculation}

We used {\tt FeynArts} \cite{Hahn00} for the generation of the diagrams. The amplitudes were
simplified using {\tt FormCalc} \cite{Hahn98}. The results in terms of Weyl-spinor chains and
coefficients containing the tensor one-loop integrals have been translated to dedicated {\tt C++}
code for the numerical evaluation. The calculation has been done in the 't Hooft--Feynman gauge.

The one-loop integrals were numerically reduced to a set of standard integrals. The 5-point integrals
were written in terms of 4-point functions following Ref. \cite{Denn02} avoiding inverse Gram
determinants. The remaining tensor coefficients of the one-loop integrals were recursively reduced to
scalar integrals with the Passarino-Veltman algorithm for non-exceptional phase-space points. In the
exceptional phase-space regions the reduction of the 3- and 4-point tensor integrals was performed
using the methods of Ref. \cite{Denn05}.

The phase-space integration was performed with Monte Carlo techniques using the VEGAS algorithm.

We checked the correctness of our results on the one hand by confirming their gauge invariance and
on the other hand by performing a major part of the calculation completely within the {\tt FeynArts}/{\tt
FormCalc}/{\tt LoopTools} toolchain and comparing both results. Gauge invariance was successfully
checked and complete agreement between the two calculations was found.

\section{Results}

\begin{figure}[htb]
	\includegraphics[width=7.4cm, height=5cm]{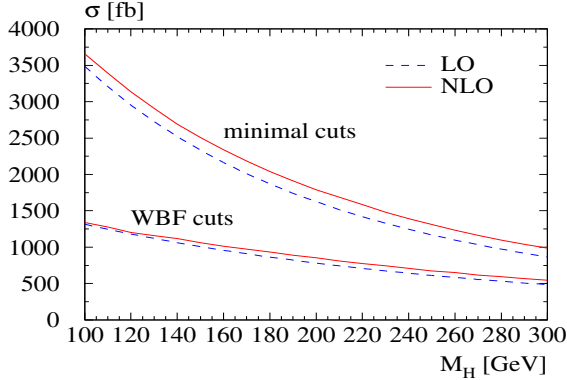}%
	\vspace*{-5mm}
	\caption{LO and NLO WBF cross section as a function of the Higgs mass $m_H$ calculated using
	\cite{Baeh08}. Minimal cuts and WBF cuts have been applied.}
	\label{fig:totcs}
\end{figure}

Fig.\ref{fig:totcs} shows the LO and NLO WBF cross sections at the LHC for comparison.
For well defined measurements several cuts have to be applied. A minimal set of cuts to ensure two
well-separated jets in the final state are given by
\begin{equation}
p_{T_j} > 20 \mbox{GeV}\,,\quad
|\eta_j|< 5\,, \quad
R > 0.6\,,
\end{equation}
where $p_{T_j}$ and $\eta_j$ are the transverse momenta and the pseudo-rapidities of the final state
jets and
\begin{equation}
R = \sqrt{ (\Delta\eta)^2 + (\Delta\phi)^2 }
\end{equation}
with $\Delta\eta = \eta_1 - \eta_2$ and $\Delta\phi = \phi_1 - \phi_2$ is the separation of the jets
in the pseudo-rapidity -- azimuthal angle plane.

The signal-to-background ratio for weak boson fusion can be improved a lot with additional cuts.
These so called WBF cuts require that two jets are well separated, reside in opposite detector
hemispheres and have a large dijet invariant mass:
\begin{equation}
|\Delta\eta| > 4.2\,,\quad 
\eta_1 \cdot \eta_2 < 0\,,\quad
m_{jj} > 600 \mbox{GeV}\,.
\end{equation}

\begin{figure*}[htb]
	\begin{center}
		\begin{tabular}{ccc}
			\includegraphics[width=5cm]{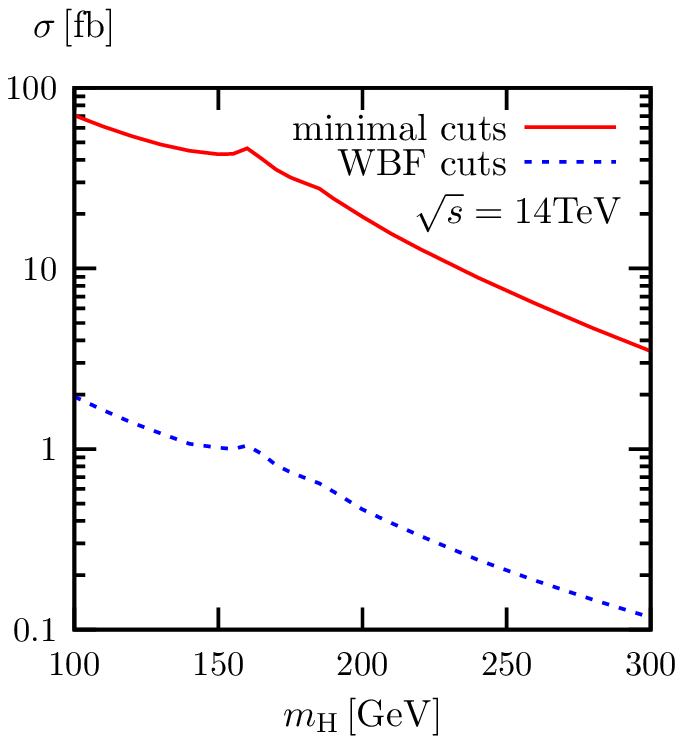} & \hspace*{0.4cm} &
			\includegraphics[width=5cm]{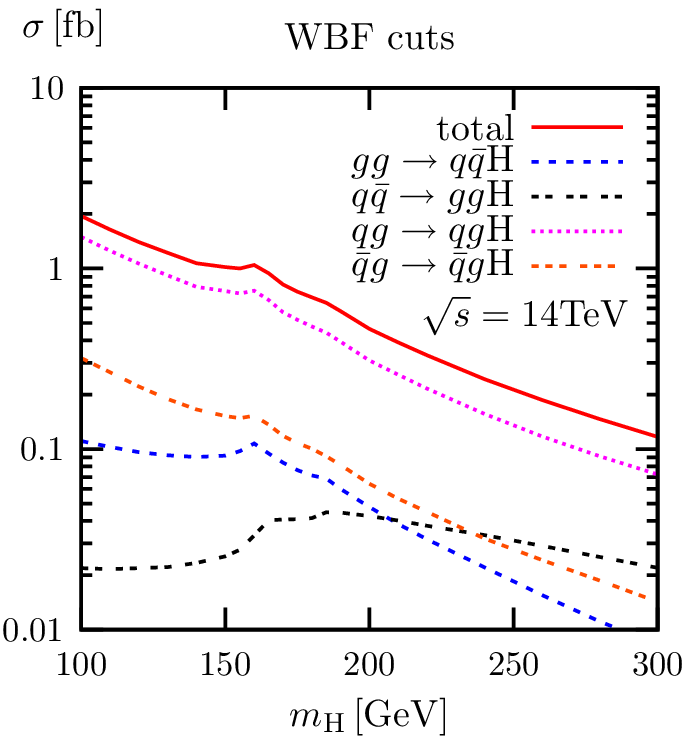} \\
			(a) & & (b) \\
			\\
			\includegraphics[width=4.4cm]{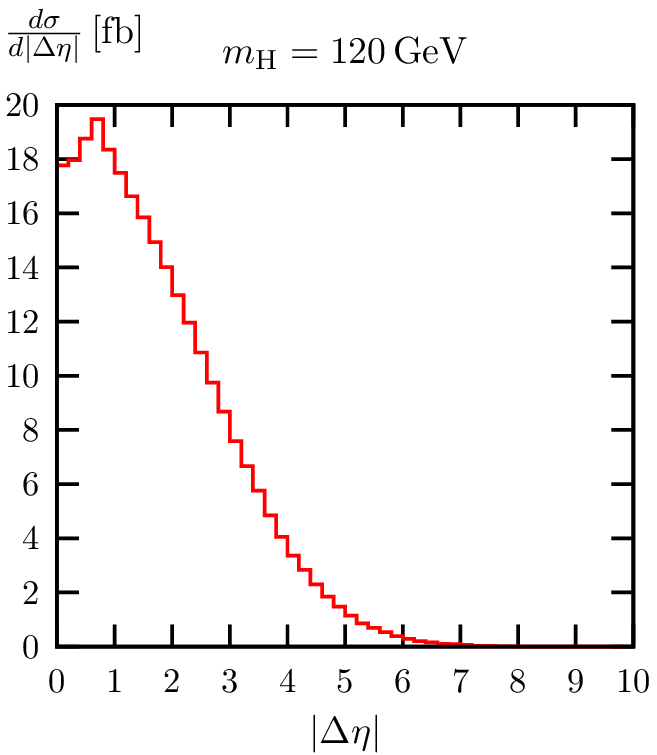} & &
			\includegraphics[width=4.95cm]{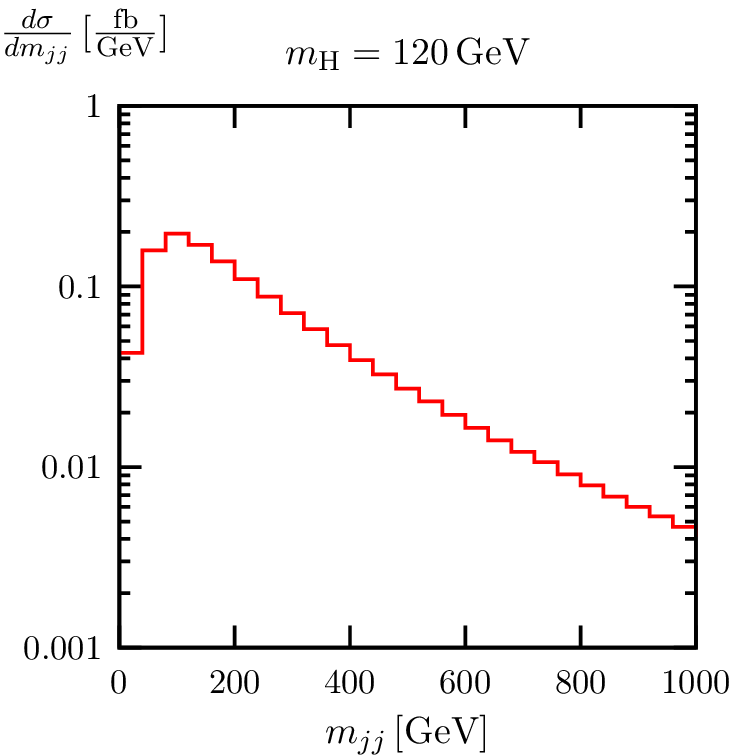} \\
			(c) & & (d)
		\end{tabular}
		\vspace*{-6mm}
		\caption[]{\label{fig:nnlo}\sloppy
		(a) NNLO contribution for the WBF cross section.
		(b) Contributions for individual initial states with WBF cuts.
		(c) Pseudo-rapidity gap $\Delta\eta$ distribution.
		(d) Invariant jet mass $m_{jj}$ distribution. }
	\end{center}
\end{figure*}

Our results for the NNLO cross section are shown in Fig.\ref{fig:nnlo}(a) as a function of the Higgs
mass. The cross section falls off very steeply for higher Higgs masses. At $m_H = 120$ GeV this
amounts to 2\% of the LO cross section.

The effect of the WBF cuts is a strong suppression of the cross section by roughly a factor of 30.
This strong suppression is in contrast to the LO and NLO WBF rates which show only a suppression by
a factor of 2--3. As the WBF cuts are designed to take advantage of the characteristic signature of
weak boson fusion, this indicates that the kinematics of the contribution here is rather different
from that of the normal WBF kinematics.

In Fig.\ref{fig:nnlo}(b) the contributions of the separate processes are shown. The largest
contribution comes from the process $qg \rightarrow qgH$ while all other processes are at least a
factor of 3 smaller.

In order to understand the strong suppression by the WBF cuts better we looked at the
distributions of the pseudo-rapidity separation $\Delta\eta$ and the invariant mass $m_{jj}$.
Fig.\ref{fig:nnlo}(c) and Fig.\ref{fig:nnlo}(d) show the distributions for a Higgs mass $m_H=120$
GeV. Especially the $\Delta\eta$ distribution explains the strong suppression.

\section{Summary}

We presented a calculation that allows for an estimation of the order $\alpha^3\alpha_s^2$
contributions to the Higgs production in the weak boson fusion channel. A possible deterioration of
this important channel for the Higgs discoveries at the LHC can be ruled out by this calculation
due to the small remaining cross section after the weak boson cuts.


\begin{thebibliography}{9}
\bibitem{Rain97} D.~L.~Rainwater and D.~Zeppenfeld,
  JHEP {\bf 9712} (1997) 005 [arXiv:hep-ph/9712271].
\bibitem{Han92} T.~Han, G.~Valencia and S.~Willenbrock,
  Phys.\ Rev.\ Lett.\  {\bf 69} (1992) 3274 [arXiv:hep-ph/9206246].
\bibitem{Djou99} A.~Djouadi and M.~Spira,
  Phys.\ Rev.\  D {\bf 62} (2000) 014004 [arXiv:hep-ph/9912476].
\bibitem{Harl08} R.~V.~Harlander, J.~Vollinga and M.~M.~Weber,
	Phys.\ Rev.\  D {\bf 77} (2008) 053010 [arXiv:0801.3355 [hep-ph]].
\bibitem{Zijl92} E.~B.~Zijlstra and W.~L.~van Neerven,
  Nucl.\ Phys.\  B {\bf 383} (1992) 525.
\bibitem{Hahn00}
  T.~Hahn,
  Comput.\ Phys.\ Commun.\  {\bf 140} (2001) 418 [arXiv:hep-ph/0012260].
\bibitem{Hahn98}
  T.~Hahn and M.~Perez-Victoria,
  Comput.\ Phys.\ Commun.\  {\bf 118} (1999) 153 [arXiv:hep-ph/9807565].
\bibitem{Denn02}
  A.~Denner and S.~Dittmaier,
  Nucl.\ Phys.\  B {\bf 658} (2003) 175 [arXiv:hep-ph/0212259].
\bibitem{Denn05}
  A.~Denner and S.~Dittmaier,
  Nucl.\ Phys.\  B {\bf 734} (2006) 62 [arXiv:hep-ph/0509141].
\bibitem{Baeh08} M.~B\"ahr {\it et al.},
  {\it VBFNLO --- NLO parton level Monte Carlo for Vector Boson Fusion},
  {\tt http://www-itp.particle.uni- karlsruhe.de/\~\/vbfnloweb/}
\end{thebibliography}
\end{document}